\documentclass[12pt,english]{article}
\pdfoutput=1
\usepackage{lmodern}

\usepackage[T1]{fontenc}
\usepackage[latin9]{inputenc}
\usepackage{geometry}
\usepackage{color}
\usepackage{babel}
\usepackage{mathtools}
\usepackage{amsmath}
\usepackage{amssymb}
\usepackage{graphicx}
\usepackage{esint}
\usepackage[numbers,sort&compress]{natbib}
\usepackage[unicode=true,pdfusetitle,
 bookmarks=true,bookmarksnumbered=false,bookmarksopen=false,
 breaklinks=false,pdfborder={0 0 1},backref=false,colorlinks=true]
 {hyperref}
\hypersetup{
 citecolor=black,linkcolor=black,urlcolor=black}
\makeatletter

 \@ifundefined{textcolor}{}
 {
   \definecolor{BLACK}{gray}{0}
   \definecolor{WHITE}{gray}{1}
   \definecolor{RED}{rgb}{1,0,0}
   \definecolor{GREEN}{rgb}{0,1,0}
   \definecolor{BLUE}{rgb}{0,0,1}
   \definecolor{CYAN}{cmyk}{1,0,0,0}
   \definecolor{MAGENTA}{cmyk}{0,1,0,0}
   \definecolor{YELLOW}{cmyk}{0,0,1,0}
 }
\usepackage{babel}
\makeatletter
\def\simgt{\mathrel{\lower2.5pt\vbox{\lineskip=0pt\baselineskip=0pt
           \hbox{$>$}\hbox{$\sim$}}}}
\def\simlt{\mathrel{\lower2.5pt\vbox{\lineskip=0pt\baselineskip=0pt
           \hbox{$<$}\hbox{$\sim$}}}}
\makeatother

\newcommand{\be}{\begin{equation}}
\newcommand{\ee}{\end{equation}}
\newcommand{\bea}{\begin{eqnarray}}
\newcommand{\eea}{\end{eqnarray}}
\newcommand{\eq}[1]{\begin{align}\begin{split}#1\end{split}\end{align}}
\newcommand{\Ref}[1]{Ref.~\cite{#1}}

\newcommand{\Eq}[1]{Eq.~\eqref{#1}}
\newcommand{\Eqs}[2]{Eqs.~\eqref{#1} and \eqref{#2}}
\newcommand{\Sec}[1]{Sec.~\ref{#1}}

\renewcommand{\AA}{{\cal A}}

\newcommand{\EE}{{\cal E}}
\newcommand{\FF}{{\cal F}}
\newcommand{\PP}{{\cal P}}
\newcommand{\HH}{{\cal H}}
\newcommand{\LL}{{\cal L}}
\newcommand{\MM}{{\cal M }}
\newcommand{\NN}{{\cal N }}
\newcommand{\RR}{{\cal R }}
\renewcommand{\SS}{{\cal S }}
\newcommand{\TT}{{\cal T }}
\newcommand{\XX}{{\cal X }}
\newcommand{\YY}{{\cal Y }}
\newcommand{\ZZ}{{\cal Z }}

\makeatletter
\newcommand{\subalign}[1]{%
  \vcenter{%
    \Let@ \restore@math@cr \default@tag
    \baselineskip\fontdimen10 \scriptfont\tw@
    \advance\baselineskip\fontdimen12 \scriptfont\tw@
    \lineskip\thr@@\fontdimen8 \scriptfont\thr@@
    \lineskiplimit\lineskip
    \ialign{\hfil$\m@th\scriptstyle##$&$\m@th\scriptstyle{}##$\crcr
      #1\crcr
    }%
  }
  }
  \makeatother

\begin{document}
	\interfootnotelinepenalty=10000
	\baselineskip=18pt
		\hfill CALT-TH-2017-051
	\hfill
	
	\vspace{2cm}
	\thispagestyle{empty}
	\begin{center}
		{\LARGE \bf 
Pions as Gluons in Higher Dimensions
		}
		\\
		\bigskip\vspace{1.cm}{
			{\large Clifford Cheung,${}^a$ Grant N. Remmen,${}^{b,c}$\\ \smallskip Chia-Hsien Shen,${}^{d}$ and Congkao Wen${}^{a,d}$}
		} \\[7mm]
		{\it  ${}^a$Walter Burke Institute for Theoretical Physics, \\[-1mm]
			California Institute of Technology, Pasadena, CA 91125
			\\
			${}^b$Berkeley Center for Theoretical Physics, Department of Physics,\\[-1mm]
University of California, Berkeley, CA 94720
\\
${}^c$Theoretical Physics Group, Lawrence Berkeley National Laboratory, Berkeley, CA 94720
			\\
			${}^d$Mani L. Bhaumik Institute for Theoretical Physics, \\[-1mm]
			 Department of Physics and Astronomy, UCLA, Los Angeles, CA 90095
			}\let\thefootnote\relax\footnote{e-mail: \url{clifford.cheung@caltech.edu}, \url{grant.remmen@berkeley.edu}, \url{chshen@physics.ucla.edu}, \url{cwen@caltech.edu }} \\
	\end{center}
	\bigskip
	\centerline{\large \bf  Abstract}	
	
\begin{quote} \small

We derive the nonlinear sigma model as a peculiar dimensional reduction of Yang-Mills theory. In this framework, pions are reformulated as higher-dimensional gluons arranged in a kinematic configuration that only probes cubic interactions.  This procedure yields a purely cubic action for the nonlinear sigma model that exhibits a symmetry enforcing color-kinematics duality.  Remarkably, the associated kinematic algebra originates directly from the Poincar\'e algebra in higher dimensions.   Applying the same construction to gravity yields a new quartic action for Born-Infeld theory and, applied once more, a cubic action for the special Galileon theory.  Since the nonlinear sigma model and special Galileon are subtly encoded in the cubic sectors of Yang-Mills theory and gravity, respectively, their double copy relationship is automatic.

\end{quote}

\setcounter{footnote}{0}

\newpage
\setcounter{tocdepth}{2}
\tableofcontents
	
\newpage

\section{Introduction}

Recent work \cite{unified} has demonstrated how gravity encodes a unified description of Yang-Mills (YM) theory, the nonlinear sigma model (NLSM), Born-Infeld (BI) theory, and the special Galileon (SG) theory~\cite{Cheung:2014dqa,Cachazo:2014xea,Hinterbichler:2015pqa}, as originally anticipated in the context of the Cachazo-He-Yuan formalism~\cite{Cachazo:2013hca,Cachazo:2013iea,Cachazo:2014xea}.
In particular, the tree-level S-matrices of these theories can be ``transmuted'' from that of gravity via simple operators that act as differentials on the space of kinematic invariants. 

In this paper, we argue that the amplitudes construction derived in \Ref{unified} is equivalent to a peculiar version of dimensional reduction and can be implemented at the level of the action.  Physically, our construction recasts pions as gluons in a special kinematic configuration in higher dimensions, thus reformulating the NLSM in $d$ dimensions as a particular dimensional reduction of YM theory in $2d+1$ dimensions.   The resulting description coincides precisely with one recently proposed in \Ref{XYZ}, where the NLSM action is comprised purely of cubic interactions exhibiting an explicit symmetry that maintains color-kinematics duality~\cite{BCJ}.

Furthermore, by applying our dimensional reduction to gravity in $2d+1$ dimensions, we obtain a new action for BI theory in $d$ dimensions.  In this representation, the interaction vertices truncate at quartic order.  Applying this operation again to BI then yields the cubic double copy action for SG proposed in \Ref{XYZ}, which is term-by-term the square of the NLSM action previously mentioned. 

As our dimensional reduction effectively projects out all quartic interactions in the NLSM, pion scattering originates entirely from the cubic topologies of gluon scattering.  This effect offers some insight into the physical origins of double copy relations~\cite{BCJ,Bern:2010ue,Bern:2010yg}.  Since the cubic sector of gravity is trivially the square of that of YM theory, the double copy relationship is inherited by the SG and NLSM. This is reminiscent of the manifestation of the double copy in self-dual YM and gravity~\cite{Monteiro:2011pc}, but applicable in general spacetime dimension. Remarkably, by deriving the NLSM action in \Ref{XYZ} directly from YM theory, we learn that the associated kinematic algebra is actually a direct descendant of the higher-dimensional Poincar\'e algebra.

While these new actions manifest the hidden relations first found in tree-level amplitudes, they display some unconventional traits that differentiate them from the standard action formulations of the quantum field theories we consider.  In particular, these actions are typically taken to be functions of a single physical field, so properties like Bose symmetry and S-matrix unitarity are obvious. However, as discovered in \Ref{XYZ}, the new NLSM action that makes the double copy relationship explicit involves more than one type of field: there are additional auxiliary fields present that obscure the underlying Bose symmetry and S-matrix unitarity (e.g., tree-level factorization). The usual NLSM tree amplitudes are reproduced as a specific choice of external states in this new formulation. The auxiliary fields in our actions---and amplitudes going beyond this prescribed choice of external states---do not have any clear physical significance. Accordingly, the action representations we derive in this paper are physical in the sense that they reproduce the correct tree-level scattering amplitudes when our prescribed choices of external states are made.  

The construction of alternative tree-level representations of quantum field theories with auxiliary states has helped in understanding the double copy and simplifying the perturbation theory~\cite{XYZ,twofold}. As in the case of the double copy itself, the question of whether this construction extends to loop order is nontrivial and will likely involve the introduction of ghost fields, so we leave this question for future work.  When restricted to the external states that are relevant to pion scattering,  \Ref{unified} proved that both properties are present at the level of amplitudes using on-shell recursion relations~\cite{Britto:2004ap,Britto:2005fq,Cheung:2015ota}, though a more direct physical understanding is still missing. In the present paper, we design the special type of dimensional reduction precisely to realize the transmutation in \Ref{unified}, so permutation invariance and unitarity follow from the proof therein.

The remainder of this paper is organized as follows.  In \Sec{sec:amp}, we summarize the results of \Ref{unified}, which defined a set of unifying relations connecting scattering amplitudes across a spectrum of theories.  We then discuss the action-level representation of this operation for the NLSM in \Sec{sec:YM}, followed by its implications for color-kinematics duality.  Finally, we apply this construction to the gravity action to derive BI theory and the SG in \Sec{sec:gravity} and conclude in \Sec{sec:conclusions}.

\section{Amplitudes Preamble}
\label{sec:amp}

In this section, we review the mechanics of transmutation at the level of scattering amplitudes \cite{unified} and show how it is equivalent to a certain implementation of dimensional reduction.

\subsection{Unifying Relations for Amplitudes}

Consider a tree-level color-ordered scattering amplitude in YM theory. As proven in \Ref{unified}, gluons can be transmuted into pions via a simple differential operation,
\be 
\frac{\partial}{\partial (e_1 e_n)} \prod_{i=2}^{n-1} \left(  \sum_{j\neq i} p_i p_j \frac{\partial}{\partial (p_j e_i)}  \right)A(g_1,\cdots, g_n) = A(\pi_1, \cdots, \pi_n), \label{eq:pion_from_gluon}
\ee
where $p_i p_j$, $p_i e_j$, and $e_i e_j$ are Lorentz invariant products of the momenta and polarization vectors and the YM color structure on the left-hand side is mapped to the NLSM flavor structure on the right-hand side.   As required by little group covariance, the transmutation operator effectively strips off all polarization vectors in order to generate an amplitude of scalars.
The very same transmutation operator also converts tree-level amplitudes of BI photons into those of SG scalars,
\be 
\frac{\partial}{\partial (e_1 e_n)} \prod_{i=2}^{n-1} \left(  \sum_{j\neq i} p_i p_j \frac{\partial}{\partial (p_j e_i)}  \right)  A(\gamma_1,\cdots, \gamma_n) = A(\phi_1, \cdots, \phi_n). \label{eq:Gal_from_BI}
\ee
Crucially, \Eqs{eq:pion_from_gluon}{eq:Gal_from_BI} apply to amplitudes in any representation, provided they are written as a function of kinematic invariants in general spacetime dimension.  This is possible because the transmutation operators are precisely engineered to be invariant under reshuffling of terms via total momentum conservation and on-shell conditions \cite{unified}.  

Note that the right-hand sides of \Eqs{eq:pion_from_gluon}{eq:Gal_from_BI} are manifestly cyclic and permutation invariant, respectively, while the left-hand sides are not. This feature is generic: while transmutation selects two special legs, chosen here to be 1 and $n$, the final answer is independent of this choice.  As we will see, the absence of manifest cyclic and permutation invariance will persist at the action level.

In \Ref{unified} it was shown how transmutation also applies to gravity---or more precisely, the low-energy effective field theory of the closed string, which describes gravity coupled to a dilaton and two-form gauge field.  Throughout, we will for brevity refer to this multiplet of states collectively as the ``extended graviton.''\footnote{The theory of gravity coupled to a dilaton and a two-form gauge field has several aliases, including ``${\cal N}=0$ supergravity'' and the theory of the ``fat graviton.'' \cite{Luna:2016hge}}
The extended graviton amplitudes are a function of non-symmetric tensor polarizations, $e_{\mu \overline\nu}= e_{\mu} \overline e_{\overline \nu}$, and are the natural output of various ``gravity = gauge$^2$'' relations arising from the BCJ~\cite{BCJ} and KLT~\cite{KLT} constructions.  Transmuting the extended graviton amplitude yields the scattering amplitude of BI photons,
\be
\frac{\partial}{\partial (\overline e_1 \overline e_n)} \prod_{i=2}^{n-1} \left(  \sum_{j\neq i} p_i p_j \frac{\partial}{\partial (p_j \overline e_i)}  \right) A({\cal H}_1,\cdots, {\cal H}_n) = A(\gamma_1, \cdots, \gamma_n).
\label{eq:BI_amps}
\ee
Here the transmutation operator only strips off the barred polarizations, so the resulting expression is still a function of the unbarred polarizations labeling the external BI photons. Combined with \Eq{eq:Gal_from_BI}, \Eq{eq:BI_amps} shows that applying the transmutation twice to an extended graviton amplitude leads to that of SG.

\subsection{Transmutation as Special Kinematics}

The transmutation procedure outlined above is actually equivalent to a certain variation of dimensional reduction.   To understand why, we first examine the case of pions transmuted from gluons, as described in \Eq{eq:pion_from_gluon}.  With the benefit of hindsight, let us define a theory of $(2d+1)$-dimensional gluons dimensionally reduced to a $d$-dimensional subspace on which the external momenta have support.  The $(2d+1)$-dimensional momentum vector for a massless gluon is
\be 
\PP_i^\MM = (p_i^\mu ,0,0), \label{eq:dimredmom}
\ee
expressed in block form where the first and third entries are $d$-dimensional and the middle entry is one-dimensional.  Throughout, we use calligraphic indices to label the full $(2d+1)$-dimensional space and, Greek indices to label both sets of $d$-dimensional spaces. It is important to point out that this latter choice of indices is simply a convenient abuse of notation; we do not identify the two $d$-dimensional spaces.

By inspection, we see that \Eq{eq:pion_from_gluon} is equivalent to the following choice of external polarizations,
\be 
\EE_1^\MM =\EE_n^\MM   = (0,1,0) \qquad \textrm{and} \qquad 
\EE_i^\MM  = (p_i^\mu, 0, i\, p_i^\mu) \qquad \textrm{for} \qquad  i\neq 1,n. \label{eq:explicit_vectors}
\ee
This is merely a choice of polarization and the two $d$-dimensional spaces remain independent spacetime directions. In order to verify this claim it suffices to compute the kinematic invariants corresponding to \Eqs{eq:dimredmom}{eq:explicit_vectors}.  For example, the invariants built purely from momenta are
\be 
\PP_i \PP_j = p_i p_j .
\ee
Meanwhile, since the polarizations of legs 1 and $n$ are orthogonal to all other legs, we find that
\be 
\EE_1 \EE_n = 1 ,
\ee
while $\EE_i \EE_j=0$ for all other combinations due to crucial factors of the imaginary number $i$ in \Eq{eq:explicit_vectors}.  Finally, the invariants constructed from polarizations and momenta are
\be 
\PP_i \EE_j = p_i p_j \qquad \textrm{for} \qquad j \notin \{1,n\},
\ee
with $\PP_i \EE_1 = \PP_i \EE_n = 0$.  Hence, this choice of external kinematics implements precisely the differential operator in \Eq{eq:pion_from_gluon}.  To obtain this result, it was important that the gluon amplitude is linear in each of the polarization vectors.

The choice of kinematics in \Eqs{eq:dimredmom}{eq:explicit_vectors} describes a dimensional reduction from $2d+1$ dimensions down to $d$ dimensions.  Physically, legs 1 and $n$ are polarized in their own exclusive extra dimension, while legs 2 through $n-1$ describe polarizations residing in the $d$-dimensional subspaces that are proportional to the physical $d$-dimensional momentum.  In subsequent sections, we translate this special kinematic configuration into an operation at the level of the action.

\section{From Gluons to Pions}

\label{sec:YM}

Let us now apply the dimensional reduction described in the previous section to derive the NLSM from YM theory.  For YM theory in $2d+1$ dimensions, the Lagrangian is
\be 
{\cal L}_{\rm YM} = -\frac{1}{4}{\rm Tr}\left( \FF_{\MM\NN} \FF^{\MM\NN}\right)+{\cal L }_{\rm GF} \qquad \textrm{with} \qquad 
\FF_{\MM\NN} =\partial_\MM \AA_\NN - \partial_\NN \AA_\MM  - i\sqrt{2}\, \left[\AA_\MM, \AA_\NN\right], 
\ee
in units where the gauge coupling $g=2$ and the gluon fields $\AA_\MM = \AA_\MM^a T^a$ are adjoint-valued under a normalization convention where 
\eq{
{\rm Tr}\left(T^a T^b\right) = \delta^{ab} \qquad \textrm{and} \qquad [T^a,T^b]=i\sqrt{2} \, f^{abc}T^c.
}  
For simplicity we implement Feyman gauge by choosing
\eq{
{\cal L }_{\rm GF} = -\frac{1}{2}{\rm Tr} \left(\partial_\MM \AA^\MM \partial_\NN \AA^\NN\right),
}
so the full action is equal to
\be
\begin{aligned}
{\cal L}_{\rm YM} &= {\rm Tr}\left(
-\frac{1}{2}\partial_\MM \AA_\NN  \partial^\MM \AA^{\NN} 
+i\sqrt{2}\, \partial_\MM \AA_\NN  [\AA^{\MM}, \AA^{\NN}]
+\frac{1}{2} [\AA_\MM, \AA_\NN][\AA^\MM, \AA^\NN]
\right ).
\end{aligned}
\ee
In what follows, we prove how the YM action reduces to the NLSM action in \Ref{XYZ}  on the dimensional reduction corresponding to \Eq{eq:explicit_vectors}.

\subsection{Dimensional Reduction to the Nonlinear Sigma Model}

According to \Eq{eq:explicit_vectors}, the gluon field $\AA_\MM$ is split into the component fields
\be
\AA_\MM = \XX_\MM + \YY_\MM +\ZZ_\MM,
\ee
which without loss of generality can be parameterized by
\be 
\begin{aligned}
\XX_\MM&=\frac{1}{\sqrt{2}} (X_\mu,0, -iX_{\mu})\\
\YY_\MM&=(0,Y,0)\\
\ZZ_\MM&= \frac{1}{\sqrt{2}} (Z_\mu , 0, +iZ_{\mu}), \label{eq:XYZdef}
\end{aligned}
\ee
where $X_\mu$ and $Z_\mu$ are $d$-dimensional vectors and $Y$ is a scalar. Since we still have the same degrees of freedom, this is merely a change of basis. As previously mentioned, we are not identifying the two $d$-dimensional spaces and will be consistently contracting indices separately in the two factors.
We assume a flat $(2d+1)$-dimensional metric, which takes the block form\footnote{Note that both $d$-dimensional spacetime factors are separately in Lorentzian mostly-plus signature while the single extra dimension is spatial.  }
\be 
\eta_{\MM \NN} = \left(
\begin{array}{ccc}
\eta_{\mu\nu} & 0 &  0 \\
0 & 1&  0 \\
0 & 0 &  \eta_{\mu \nu} 
\end{array}
\right) ,
\ee
so the square of the gluon field is
\begin{equation}
	\AA_\MM \AA^\MM =  X_\mu Z^\mu +  Z_\mu X^\mu + Y^2.
\end{equation}
By construction, $\XX_\MM$ and $\ZZ_\MM$ have the form of polarizations of opposite helicity.  As a result, we obtain the useful identities
\be 
\begin{aligned}
&\XX_\MM \XX^\MM \!\!\!&= \ZZ_\MM \ZZ^\MM &= 0  \\
&\XX_\MM \YY^\MM \!\!\!&= \ZZ_\MM \YY^\MM &=0
\end{aligned}
\label{eq:nilpotent}
\ee 
and similarly for all analogous expressions involving derivatives on fields.   Finally, we note that, in accordance with \Eq{eq:dimredmom}, the fields are polarized in the full $(2d+1)$-dimensional space but only carry momentum in the first $d$ spacetime dimensions, so
\be 
\partial_\MM = (\partial_\mu,0,0).
\ee
Expanding the YM action in terms of $X_\mu$, $Y$, and $Z_\mu$, we obtain
\be 
\LL_{\rm YM} =  \LL^{(2)}_{\rm YM} + \LL^{(3)}_{\rm YM} + \LL^{(4)}_{\rm YM} ,
\ee
where the terms at each power in fields are
\be
\begin{aligned}
\LL^{(2)}_{\rm YM} & = {\rm Tr}\bigg(  X_\mu\Box Z^{\mu}   + \frac{1}{2}  Y \Box Y \bigg)  \\
\LL^{(3)}_{\rm YM} & =  i\,{\rm Tr}\bigg( \partial_\mu X_\nu \left[Z^{\mu}, Z^{\nu}\right]  + X_\mu\left[\partial_\nu Z^{\mu}, Z^{\nu}\right]     + Z^{\mu} \left[Y,\partial_{\mu} Y\right]   \bigg) + \left\{ X_\mu \leftrightarrow Z_\mu \right\} \\
\LL^{(4)}_{\rm YM} & =  {\rm Tr}\bigg( 
\left[X_\mu,Z^\nu\right] \left[Z^\mu,X_\nu\right] + \left[X_\mu,X_\nu\right] \left[Z^\mu,Z^\nu\right] + 2\left[X_\mu, Y\right] \left[Z^\mu,Y\right]
\bigg).
\label{eq:XYZ4}
\end{aligned}
\ee 
Remarkably, one can consistently drop the majority of terms in the action \eqref{eq:XYZ4} because they do not actually contribute to tree-level pion scattering.  This truncation is possible as a consequence of  two important simplifications, which we now discuss.

\subsubsection*{Weight Counting}

First of all, let us determine which interaction vertices actually enter into the tree-level Feynman diagrams for pion scattering.
According to \Eq{eq:explicit_vectors}, pion scattering corresponds to higher-dimensional gluon scattering where legs 1 and $n$ are $Y$ particles and all other states are longitudinally-polarized $Z_\mu$ particles.  In particular, the latter states all have polarizations proportional to their respective momenta $p_\mu$.  At the level of the amplitude, we have 
\be 
A(\pi_1, \pi_2, \ldots , \pi_{n-1}, \pi_n) = A(Y_1,  Z_{2}, \ldots,  Z_{n-1}, Y_n),
\label{eq:NLSM_amp}
\ee
which incidentally matches the prescription proposed in \Ref{XYZ}. 

As it turns out, since the external states are restricted to longitudinal $Z_\mu$ states and a pair of $Y$ states, this severely limits which interactions can contribute to the amplitude. By drawing tree-level Feynman diagrams explicitly, it becomes obvious that none of the quartic interactions can appear. Since we are interested in the NLSM, it is desirable to further simplify the action in order to make the color-kinematics duality manifest, as in \Ref{XYZ}.

To systematically enumerate which interactions in \Eq{eq:XYZ4} can appear in a tree-level scattering amplitude for the NLSM according to the external states specified by \Eq{eq:NLSM_amp}, we define a ``pseudo-helicity'' for each external particle type,
\be 
h[X_\mu]=-1, \qquad h[Y] =0,\qquad h[Z_\mu]=+1.
\ee
In analogy with helicity in four dimensions, it is natural to define the holomorphic weight of an operator, $w=n-h$, where $n$ is the total number of particles in the operator and $h$ is the sum of all pseudo-helicities in the operator.  At tree level, there is a simple addition rule for the weights.  This is because the weights satisfy $w[A] = w[A_L] + w[A_R] -2$ for an amplitude on its factorization channel, $A \sim A_L A_R$.
 For an in-depth discussion of weight counting in general, see \Ref{holomorphy}. 
 The weight of each component field is
\be 
w[X_\mu] = +2, \qquad
w[Y] = +1,\qquad
w[Z_\mu] = 0, \label{eq:field_weight}
\ee
so each term in the action in \Eq{eq:XYZ4} has weight
\be 
w \! \left[\LL^{(2)}_{\rm YM}\right] =+2,\qquad w \! \left[\LL^{(3)}_{\rm YM}\right] = +2, \,+4 , \qquad w \!  \left[\LL^{(4)}_{\rm YM}\right] = +4.
\ee
We thus learn that every term in the Lagrangian has weight $w\geq +2$.  However, since the pion scattering amplitude contains all $Z_\mu$ states except for a pair of $Y$ states, the target amplitude has weight $w=+2$.  
This implies that pion amplitudes only receive contributions from $w=+2$ interactions, so it is consistent to entirely drop all terms in the Lagrangian with weight $w >+2$. The resulting truncated action is \Eq{eq:XYZ4} with all the quartic terms and half the cubic terms dropped,
\be 
\LL_{\rm NLSM}  =  {\rm Tr}\bigg(  X_\mu\Box Z^{\mu}   + \frac{1}{2}  Y \Box Y + i\,\big( \partial_\mu X_\nu\left[Z^{\mu}, Z^{\nu}\right]  + X_\mu  \left[\partial_\nu Z^{\mu}, Z^{\nu}\right] + Z^{\mu} \left[Y,\partial_{\mu} Y\right] \big)  \bigg).
\label{eq:LNLSM}
\ee
This action is similar but not yet equal to the NLSM action proposed in \Ref{XYZ}.  

\subsubsection*{Transverse Condition}

To establish complete equivalence requires a second simplification of the action that arises from certain transverse properties of the fields.
First, we rewrite \Eq{eq:LNLSM}, up to total derivatives, as 
\be 
\LL_{\rm NLSM}  =  {\rm Tr}\bigg(  X_\mu\Box Z^{\mu}   + \frac{1}{2}  Y \Box Y + i\,\big(X_{\mu \nu}  \left[Z^{\mu}, Z^{\nu}\right] 
+Z^{\mu} \left[Y,\partial_{\mu} Y\right] \big)  \bigg) + {\cal O}(\partial_\mu Z^\mu),
\label{eq:XYZ}
\ee
where we have defined the field strength for the $X_\mu$ field,
\be 
X_{\mu\nu} = \partial_\mu X_\nu - \partial_\nu X_\mu.
\ee
The action in \Eq{eq:XYZ} differs from that of \Ref{XYZ} by terms proportional to the longitudinal component, $\partial_\mu Z^\mu$.  
As we now show, these terms are always projected out of tree-level pion amplitudes and can be consistently dropped. To understand why, consider a factor of $\partial_\mu Z^\mu$ that appears in an interaction contributing to a Feynman diagram.  If the $Z_\mu$ field contracts into an external state, then it vanishes by the on-shell conditions.  On the other hand, if the $Z_\mu$ field is contracted with an internal propagator, then the off-diagonal structure of the kinetic term links this field to the $X_\mu$ field of some internal vertex.  According to \Eq{eq:XYZ}, all interaction vertices that involve $X_\mu$ are either a function of the field strength $X_{\mu\nu}$ or are proportional to $\partial_\mu Z^\mu$.  In the former case, the field strength $X_{\mu\nu}$ simply zeroes out this longitudinal contribution.  In the latter, the internal vertex is also proportional to the longitudinal component $\partial_\mu Z^\mu$, so we can then apply the same logic from the beginning.  Thus, all factors of $\partial_\mu Z^\mu$ ultimately terminate at an external leg or on an $X_{\mu\nu}$ field strength.  Because these contributions vanish, all factors of $\partial_\mu Z^\mu$ can be consistently dropped from the action,  thus establishing the equivalence of \Eq{eq:XYZ} with the result of \Ref{XYZ},
which was originally derived from scattering amplitudes rather than dimensional reduction.

In terms of Feynman diagrams, the perturbation expansion for the action in \Eq{eq:XYZ}  is drastically simpler than that of the conventional representation of the NLSM action,
\be 
\LL_{\rm NLSM} = -\frac{ f_\pi^2}{2} {\rm Tr}\left[ \partial^\mu U^{-1}\partial_\mu U\right],
\ee
where $U = \exp(i \pi^a T^a/f_\pi)$ and $f_\pi$ is the pion decay constant.   The exponential form of the nonlinear field generates an infinite tower of higher- and higher-order interactions that contribute unnecessary complexity to the Feynman diagrammatic expansion, and obscures the color-kinematics duality in NLSM~\cite{Chen:2013fya,Du:2016tbc,Carrasco:2016ldy,Carrasco:2016ygv,Du:2017kpo,Chiodaroli:2017ngp}. In contrast, the NLSM representation in \Eq{eq:XYZ} is purely cubic and manifests the color-kinematics duality inherits from YM.  Note that the pion decay constant is absorbed into the normalization of the longitudinal polarizations of the $Z_\mu$ external states.  

\subsection{Color-Ordered Formulation}
For future reference, we summarize here the color-ordered Feynman rules derived from the NLSM action in \Eq{eq:XYZ}.
Since we are in Feynman gauge, the propagators take the simple form,
\be 
\begin{aligned}
	\begin{gathered}\includegraphics[scale=0.3]{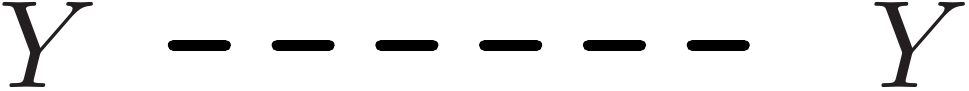}\end{gathered}\quad &= \quad -\frac{i}{p^2}  \\
	\begin{gathered}\includegraphics[scale=0.33]{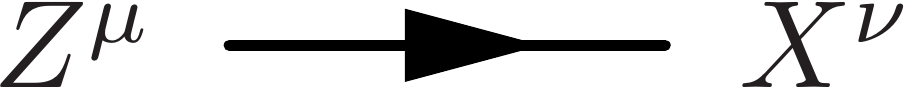}\end{gathered}\quad &= \quad  -\frac{i}{p^2}\eta^{\mu\nu},
\end{aligned}
\ee
where the $X_\mu$ and $Z_\mu$ fields are conjugate particles. The three-particle Feynman vertices are 
\be 
\begin{aligned}
	\begin{gathered}
	\includegraphics[scale=0.3]{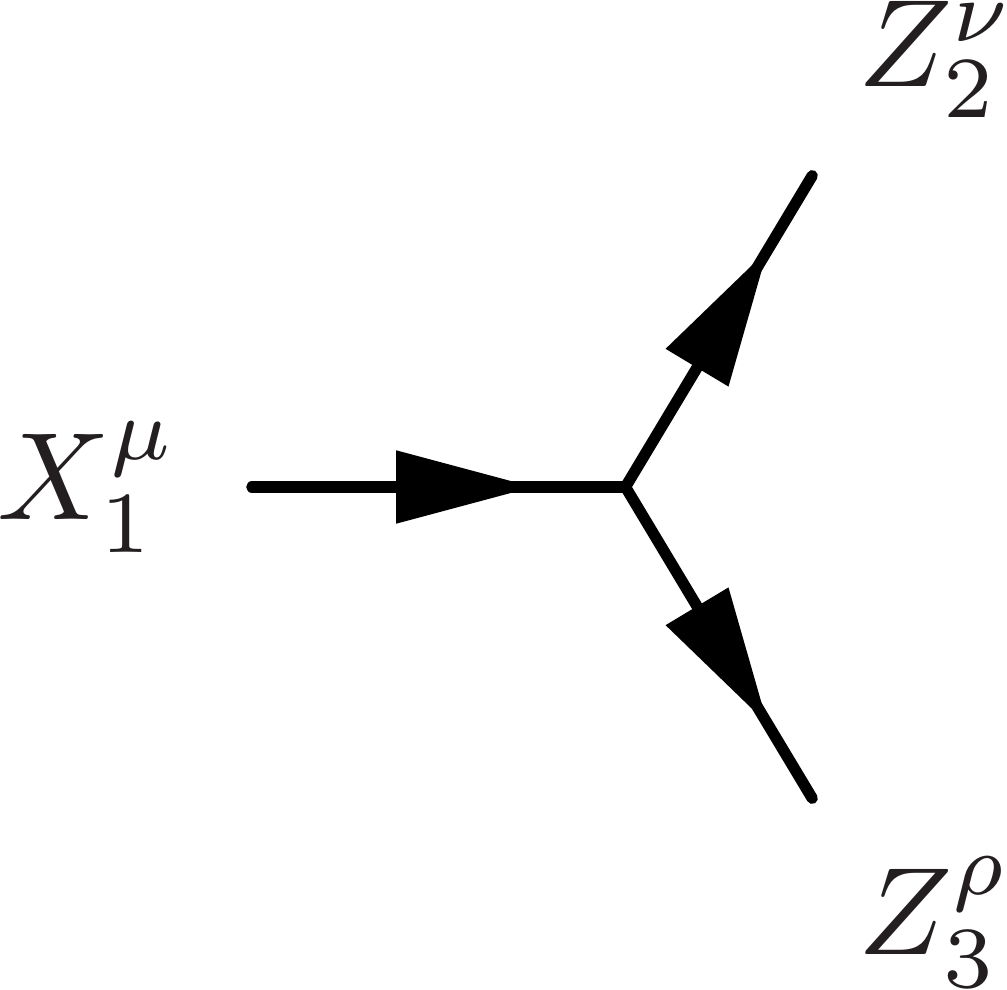}
	\end{gathered}\quad &= \quad 2i\,(p_{1}^{\nu}\eta^{\rho\mu} - p_1^\rho \eta^{\mu\nu})  \\ \; \\
	\begin{gathered} \includegraphics[scale=0.3]{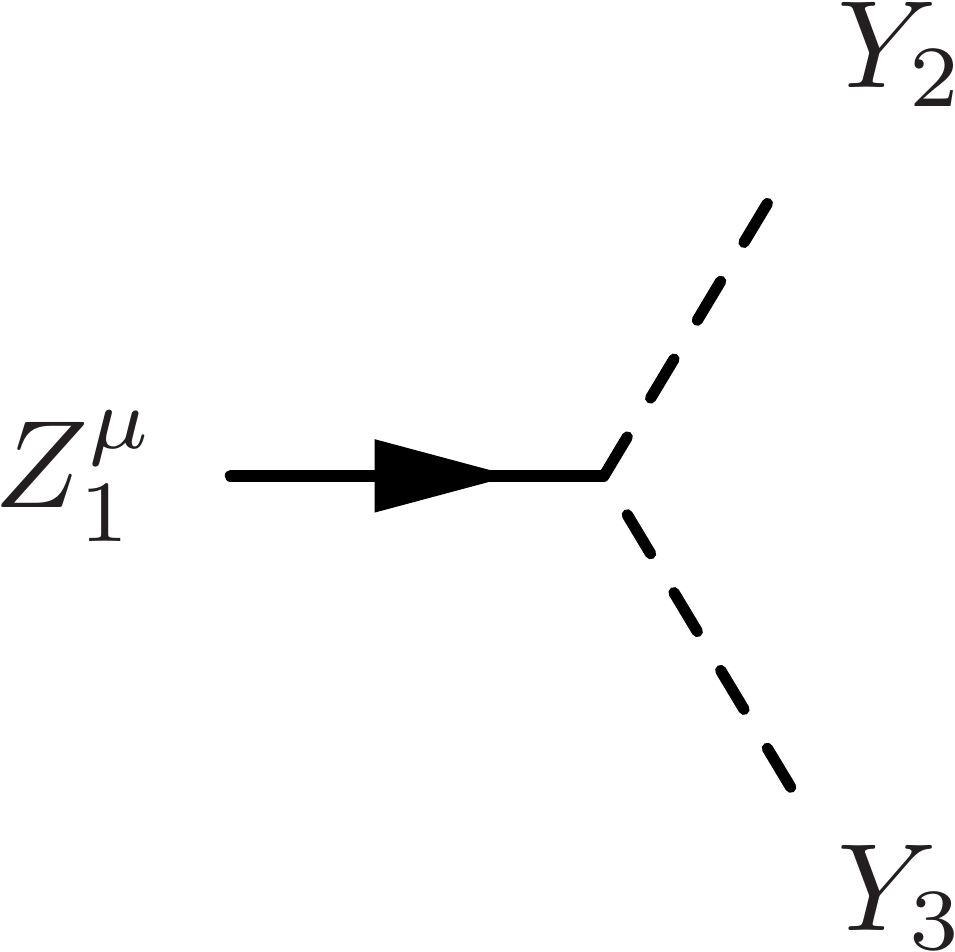} \end{gathered}\quad &= \quad  -i\,(p_2^\mu-p_3^\mu),
\end{aligned}
\ee
which are far simpler to implement than Feynman rules in the conventional approach to perturbation theory in the NLSM~\cite{Kampf:2012fn,Kampf:2013vha}.
Recall the NLSM amplitude is given by the states chosen in Eq.~\eqref{eq:NLSM_amp}.

Note that the color-ordered formulation naturally arises from YM action in the Gervais-Neveu gauge \cite{Gervais:1972tr},
\be 
	{\cal L}_{\rm YM} = {\rm Tr}\left( -\frac{1}{2} \partial_\MM \AA_\NN \partial^\MM \AA^\NN - 2\sqrt{2}i\,  \partial_\MM \AA_\NN \AA^\MM   \AA^\NN+ \, \AA_\MM \AA_\NN \AA^\MM \AA^\NN\right),
\ee
which is dimensionally reduced to 
\be 
	\LL_{\rm NLSM}  =  {\rm Tr}\bigg(  X_\mu\Box  Z^\mu  + \frac{1}{2}  Y \Box Y + 2i\, X_{\mu\nu} Z^\mu Z^\nu    + 2i\,Z^\mu  Y \partial_\mu Y \bigg),
	\label{eq:XYZ_GN}
\ee
up to terms that may be consistently dropped as a consequence of weight counting or the transverse condition discussed previously.

For the sake of completeness, we also remind the readers that the tree-level pion amplitudes are reproduced by the amplitudes in Eq.~\eqref{eq:NLSM_amp}. According to the special kinematics in Eq.~\eqref{eq:explicit_vectors}, the $d$-dimensional polarizations for $Z$ particles are chosen as the longitudinal mode, $\epsilon^{\mu}_{Z} = p^\mu$. Note that the choice of longitudinal polarization does not contradict with the transverse condition discussed earlier. The transverse condition applies to the irrelevance of interactions proportional to $\partial_{\mu} Z^{\mu}$ which has nothing to do with the choice of polarization $\epsilon^{\mu}_{Z}$.

\subsection{Kinematic Algebra as Poincar\'e Algebra}

As emphasized in \Ref{XYZ}, the Feynman diagrams associated with the NLSM action in \Eq{eq:LNLSM} automatically satisfy the Jacobi identities and are thus manifestly compliant with color-kinematics duality.  Remarkably, the Jacobi identities are enforced by a symmetry of the NLSM action, 
\be 
\begin{aligned}
 \delta_{X} \! \left( 
\begin{array}{c}
X_\mu \\
Y \\
Z_\mu
\end{array}
\right) &= \left( 
\begin{array}{c}
\theta_{X\mu \nu} Z^\nu\\
0\\
0
\end{array}
\right)  \\
 \delta_{Y} \! \left( 
\begin{array}{c}
X_\mu \\
Y \\
Z_\mu
\end{array}
\right) &= \left( 
\begin{array}{c}
\theta_{Y\mu} Y\\
-\theta_{Y\mu} Z^\mu\\
0
\end{array}
\right) \\
 \delta_Z \! \left( 
\begin{array}{c}
X_\mu \\
Y \\
Z_\mu
\end{array}
\right) &= \left( 
\begin{array}{c}
\theta_Z^{\nu} \partial_\nu X_\mu +\partial_\mu \theta_Z^{\nu}X_\nu\\
\theta_Z^{\nu} \partial_\nu Y \\
\theta_Z^{\nu} \partial_\nu Z_\mu -\partial_\nu \theta_{Z\mu} Z^\nu
\end{array}
\right) ,
\label{eq:XYZ_symmetries}
\end{aligned}
\ee
where $\theta_{X\mu\nu} = \partial_\mu \theta_{X\nu} - \partial_\nu \theta_{X\mu}$.  In particular, the Noether current conservation equations for these symmetries are literally equal to the Jacobi identities for kinematic numerators, modulo terms that vanish under the transverse conditions discussed earlier.

As noted in \Ref{XYZ}, the $\delta_Z$ transformations are simply Poincar\'e transformations acting on the $d$-dimensional subspace.  This is obvious if we identify $ \theta_Z^\mu = a^\mu + b^{\mu\nu}x_\nu$, where $a$ is a constant vector labeling translations and $b$ is a constant antisymmetric matrix labeling rotations and boosts.

But what of the remaining symmetries, $\delta_X$ and $\delta_Y$? By recasting the NLSM as a dimensional reduction of YM, we learn that these symmetries have a geometric origin---namely, Lorentz boosts in higher dimensions! Concretely, consider a matrix parameterizing a Lorentz transformation acting on the extra-dimensional space,
\begin{equation}
	\Lambda_{\MM\NN} =
	\left (\begin{array}{ccc}
		0 & 0  &  0  \\
		0 & 0  &  i\sqrt{2}\theta_{Y \nu} \\
		0 & -i\sqrt{2}\theta_{Y \mu}  &  -2 \theta_{X \mu\nu} \\
	\end{array}
	\right).
\end{equation} 
These transformations act rigidly on the indices of fields and do not involve derivatives because there are no momenta flowing in the extra dimensions.  The extra-dimensional Lorentz transformation shifts the gluon field by $\AA_\MM \rightarrow \AA_\MM + \tilde \delta \AA_\MM$, which in terms of the component fields is
\begin{equation}
\begin{split}
	\tilde \delta_X \! \left( 
	\begin{array}{c}
		X_\mu \\
		Y \\
		Z_\mu
	\end{array}
	\right) &= \left( 
	\begin{array}{c}
		-\theta_{X\mu \nu}(X^{\nu}-Z^\nu)\\
		0\\
		\theta_{X\mu \nu}(X^{\nu}-Z^\nu)
	\end{array}
	\right) \\
	\tilde \delta_Y \! \left( 
	\begin{array}{c}
		X_\mu \\
		Y \\
		Z_\mu
	\end{array}
	\right) &= \left( 
	\begin{array}{c}
		\theta_{Y\mu} Y\\
		\theta_{Y\mu} (X^\mu-Z^\mu)\\
		-\theta_{Y\mu} Y
	\end{array}
	\right).
	\end{split}
	\label{eq:YM_symmetries}
\end{equation}
At present, these symmetries still differ from Eq.~\eqref{eq:XYZ_symmetries}. However, as discussed earlier, the NLSM is defined by a truncated version of the YM action.  As a result of the truncation, a symmetry in YM is not guaranteed to be a symmetry of the NLSM.

Nevertheless, a close descendant of the extra-dimensional Lorentz symmetry is still preserved under weight truncation. To see why, recall that the original YM action can be partitioned into two weight sectors, $\LL_{\rm YM} = \LL_{\rm YM}^{(w=2)}+\LL_{\rm YM}^{(w=4)}$. The transformations in Eq.~\eqref{eq:YM_symmetries} shift the weights of the component fields by
\begin{equation}
w \! \left[\tilde \delta_X \right] = 0,\,\pm2 \qquad \text{and} \qquad
w \! \left[\tilde \delta_Y \right] = \pm1,
\end{equation}
so extra-dimensional Lorentz transformations mix terms of different weight. In order to determine the component of the Lorentz transformation that leaves $\LL_{\rm YM}^{(w=2)}$ invariant, we simply drop all transformations that are an invariance only with the help of $\LL_{\rm YM}^{(w=4)}$.    These transformations can never be symmetries of the truncated action.  On the other hand, for transformations that shift the weights strictly negatively, it is impossible for any variation of $\LL_{\rm YM}^{(w=4)}$ to ever cancel a variation in $\LL_{\rm YM}^{(w=2)}$ because these terms are already separated in weight.  Thus, for a negative shift in weight, $\LL_{\rm YM}^{(w=2)}$ will be itself invariant.  Truncating the symmetry transformation in \Eq{eq:YM_symmetries} down to terms that shift the weight by $-1$ and $-2$, we obtain the color-kinematics symmetry of the NLSM shown in \Eq{eq:XYZ_symmetries}.  In summary, color-kinematics duality in the NLSM arises from a higher-dimensional spacetime symmetry of YM theory. 

\section{From Gravitons to Photons and Galileons}

\label{sec:gravity}
The construction described above can be applied straightforwardly to gravity.   However, as discussed in \Ref{unified}, the natural theory to which to apply transmutation is the low-energy effective theory of the closed string.  The action, $S_{\rm G} = \int \mathrm{d}^D x \, {\cal L}_{\rm G}$, describes a metric ${g}_{\MM \NN}$ coupled to an antisymmetric two-form ${B}_{\MM \NN}$ and a dilaton $\phi$, with interactions given by
\be 
{\cal L}_{\rm G} = \sqrt{-g}\left[\frac{2}{\kappa^2} R  - \frac{1}{2(D-2)} \partial_\MM \phi  \partial^\MM \phi -\frac{1}{6} e^{-2\kappa \phi/(D-2)} \partial_{[\MM} B_{\NN\RR]} \partial^{[\MM} B^{\NN\RR]}\right] + \mathcal{L}_{\rm GF},
\ee
where $\kappa^2 = 32\pi G$. 
In the conventional picture, one expands the graviton in perturbations,
\be 
g_{\MM \NN} = \eta_{\MM\NN} + \kappa  h_{\MM\NN},
\ee
 treating $h_{\MM\NN}$, $B_{\MM\NN}$, and $\phi$ as distinct fields.  For our purposes, however, it will be convenient to repackage the degrees of freedom into a single extended graviton field described by a general tensor $\HH_{\MM \overline \NN}$.  An action of this form was derived in \Ref{twofold} in the context of pure gravity, but in fact also reproduces all extended graviton amplitudes as well.  Unfortunately,  the associated propagator deviates from the simple $1/p^2$ Feynman propagator form, so we will not consider the action of \Ref{twofold} further here.

An action-level version of transmutation requires an extended graviton action expressed in terms of ${\cal H}_{\MM \overline \NN}$ with a simple propagator going as $1/p^2$.   To derive such an action, we build a general ansatz for an effective field theory of the extended graviton $\HH_{\MM \overline \NN}$ and constrain its coefficients to match known tree-level amplitudes constructed from the KLT relations \cite{KLT}. Among the family of resulting actions, we choose the remaining free coefficients to simplify our results by reducing the number of terms in the Lagrangian. In natural units where $\kappa=1$, our resulting extended graviton action is
\be 
 \mathcal{L}_{\textrm{G}} = \mathcal{L}^{(2)}_{\textrm{G}} + \mathcal{L}^{(3)}_{\textrm{G}}+ \mathcal{L}^{(4)}_{\textrm{G}}+ \cdots,\label{eq:boxactionsum}
\ee
where the terms at each order are
\be
\begin{aligned}
\mathcal{L}^{(2)}_{\textrm{G}} =& \phantom{ +} \frac{1}{2} \HH_{\MM \overline \NN}\Box \HH^{\MM \overline \NN}\\
\mathcal{L}^{(3)}_{\textrm{G}} =& \phantom{ +} \frac{1}{2} \HH^{\MM \overline \NN} \partial_\MM \HH_{\RR \overline \SS} \partial_{\overline \NN} \HH^{\RR \overline \SS} 
+ \frac{1}{2} \HH^{\MM \overline \NN} \partial_{\overline \NN} \HH_{\MM \overline \RR} \partial_{\SS} \HH^{\SS \overline \RR}
- \HH^{\MM \overline \NN} \partial_{\SS} \HH_{\MM \overline \RR} \partial_{\overline \NN} \HH^{\SS \overline \RR} \\
\mathcal{L}^{(4)}_{\textrm{G}} =& \phantom{ +} \frac{1}{32}\HH_{\MM\overline \NN}\HH^{\MM \overline \NN} \partial_\TT \HH_{\RR \overline \SS} \partial^\TT \HH^{\RR \overline \SS} -\frac{1}{16} \HH_{\MM\overline\NN}\HH^{\MM\overline\NN} \partial_\TT \HH_{\RR\overline\SS} \partial^\RR \HH^{\TT\overline\SS} \\& 
+ \frac{1}{4} \HH^{\MM \overline\NN} \HH_{\RR\overline\NN} \partial^\RR \HH^{\TT\overline\SS} \partial_\TT \HH_{\MM\overline\SS} + \frac{1}{8} \HH^{\MM\overline\NN}\HH^{\RR\overline\SS}\partial_\TT\HH_{\RR\overline\NN} \partial^\TT \HH_{\MM\overline\SS} \\&
-\frac{1}{4}\HH^{\MM\overline\NN}\HH^{\RR\overline\SS}\partial_\RR \HH_{\TT\overline\NN} \partial^\TT \HH_{\MM\overline\SS}-\frac{1}{8}\HH^{\MM\overline\NN}\HH_{\RR\overline\SS}\partial_{\overline\NN} \HH_{\MM\overline\TT} \partial^{\overline\SS} \HH^{\RR\overline\TT}\\&
-\frac{1}{8}\HH^{\MM\overline\NN}\HH_{\MM\overline\RR}\partial_{\overline\NN} \HH_{\SS\overline\TT} \partial^{\overline\RR} \HH^{\SS\overline\TT} - \frac{1}{8}\HH^{\MM\overline\NN}\HH_{\RR\overline\NN} \partial_{\MM} \HH_{\SS\overline\TT} \partial^\RR \HH^{\SS\overline\TT}
\\&
+\frac{1}{4} \HH^{\MM\overline\NN}\HH_{\RR\overline\SS} \partial_{\overline\NN} \HH^{\RR\overline\TT} \partial^{\overline\SS} \HH_{\MM\overline\TT}. \label{eq:Sfat}
\end{aligned}
\ee
As we will explain, terms of higher order will be not be needed for our analysis.

\subsection{Dimensional Reduction to Born-Infeld Theory}
\label{sec:BI}

Next, let us implement the procedure described in the above sections to derive the BI action from the extended graviton action in  \Eqs{eq:boxactionsum}{eq:Sfat}.  We take the unbarred and barred indices in the extended graviton action to run over $d$ dimensions and $2d+1$ dimensions, respectively, so the extended graviton field is a $d\times(2d+1)$ matrix,
\be 
\HH_{\mu \overline \NN}  = \XX_{\mu \overline \NN}+\YY_{\mu \overline\NN}+\ZZ_{\mu\overline\NN},
\ee
where each component field is given by
\eq{
\XX_{\mu \overline \NN}=&\;\frac{1}{\sqrt{2}} (X_{\mu \bar \nu},0, -iX_{\mu \bar\nu})\\
\YY_{\mu \overline\NN}=&\;(0,Y_\mu,0)  \\
\ZZ_{\mu \overline\NN}=&\;\frac{1}{\sqrt{2}} (Z_{\mu\bar\nu} , 0, +iZ_{\mu\bar\nu}). \label{eq:XYZ_BI}
}
These definitions enforce a similar nilpotency condition as before,
\eq{
\XX_{\mu \overline \RR} \XX_\nu^{\; \overline \RR} &= \ZZ_{\mu \overline \RR} \ZZ_\nu^{\; \overline \RR} = 0 \\
\XX_{\mu \overline \RR} \YY_\nu^{\; \overline \RR} &=\ZZ_{\mu \overline \RR} \YY_\nu^{\; \overline \RR}  =0,
}
and likewise for terms involving derivatives.   We also note that the derivatives only have support on the $d$-dimensional subspace, so
\eq{
\partial_\MM = (\partial_\mu,0,0) \qquad \textrm{and} \qquad  \partial_{\overline \MM} = (\partial_{\bar \mu},0,0).
}
Plugging into the extended graviton action in \Eq{eq:Sfat}, we obtain a new action for the BI theory, 
\be 
\LL_{\rm BI} =  \LL^{(2)}_{\rm BI} + \LL^{(3)}_{\rm BI} + \LL^{(4)}_{\rm BI} ,
\ee
where the terms at each power are given by
\be 
\begin{aligned}
\LL^{(2)}_{\rm BI} =& \phantom{ + }   X_{\mu \bar\nu} \Box Z^{\mu\bar \nu} + \frac{1}{2} Y^\mu \Box Y_\mu \\
\LL^{(3)}_{\rm BI} =& \phantom{ + }  
\frac{1}{2\sqrt{2}} Z^{\mu \bar \nu} \partial_\mu Y_{\rho} \partial_{\bar \nu} Y^{\rho}
+ \frac{1}{2\sqrt{2}} Z^{\mu \bar \nu} \partial_{\bar\nu} Y_{\mu} \partial_{\rho} Y^{\rho} 
- \frac{1}{\sqrt{2}} Z^{\mu \bar \nu}\partial_{\rho} Y_{\mu} \partial_{\bar\nu} Y^{\rho}  \\&+ \{ Y_\mu Y_\nu \rightarrow X_{\mu \bar \rho} Z_\nu^{\;\;\bar \rho} +Z_{\mu \bar \rho} X_\nu^{\;\;\bar \rho} \} + {\cal O}(\partial_{\bar \nu}Z^{\mu\bar\nu}) \\
\LL^{(4)}_{\rm BI}  =& 
- \frac{1}{16} Z^{\mu \bar \nu} Z_\mu^{\;\; \bar \rho} \partial_{\bar \nu} Y_\sigma \partial_{\bar \rho} Y^\sigma 
- \frac{1}{16} Z^{\mu \bar \nu} Z^{\rho \bar\sigma} \partial_{\bar \nu} Y_\mu \partial_{\bar \sigma} Y_\rho
+ \frac{1}{8} Z^{\mu \bar \nu} Z^{\rho \bar\sigma} \partial_{\bar \nu} Y_\rho \partial_{\bar\sigma} Y_\mu \\&+ \{ Y_\mu Y_\nu \rightarrow X_{\mu \bar \rho} Z_\nu^{\;\;\bar \rho} +Z_{\mu \bar \rho} X_\nu^{\;\;\bar \rho}  \} + {\cal O}(\partial_{\bar \nu}Z^{\mu\bar\nu}).\label{eq:LagrBI}
 \end{aligned}
 \ee
 Here we have dropped all  interactions at quintic order and higher because they can be truncated by a weight counting argument that will be discussed shortly.  Moreover, we have separated off terms proportional to $\partial_{\bar \nu}Z^{\mu\bar\nu}$  because they can be discarded due to an analogue of the transverse conditions discussed earlier.

\subsubsection*{Weight Counting}

Our earlier weight counting arguments are straightforwardly generalized to the case of gravity. Since the dimensional reduction is only applied to the barred indices, the weights are defined in the same way as in Eq.~\eqref{eq:field_weight}. Following Eq.~\eqref{eq:BI_amps}, the tree-level BI amplitude is 
\be 
	A(\gamma_1, \gamma_2, \ldots , \gamma_{n-1}, \gamma_n) = A(Y_1,  Z_{2}, \ldots,  Z_{n-1}, Y_n),\label{eq:BIampstructure}
\ee
corresponding to a pair of $Y_{\mu}$ fields interacting with the $Z_{\mu\bar\nu}$ states that are longitudinally polarized on the barred index. Concretely, the $Y_{\mu}$ particles have the same polarization vectors $e_\mu$ of the corresponding BI photons and the $Z_{\mu\bar{\nu}}$ particles have polarization tensors proportional to  $e_\mu p_{\bar \nu}$. Note that these external states are simply the tensor product of BI photon polarizations $e_{\mu}$ with the $Y$ and $Z_{\bar{\nu}}$ external states for the NLSM in Eq.~\eqref{eq:NLSM_amp}.
Since the BI amplitude has uniform weight $w=+2$, we can truncate the action by dropping all terms with weight $w>+2$.

The extended gravity interactions take the schematic form
\eq{	
\mathcal{L}^{(3)}_{\textrm{G}} \sim  \HH^3 \partial \bar{\partial} \qquad \textrm{ and } \qquad
	\mathcal{L}^{(4)}_{\textrm{G}} \sim \HH^4  \partial \partial + \HH^4 \bar{\partial} \bar{\partial},
	\label{eq:Sfat_schematic}
}where we ignore all index structure except the barred or unbarred nature of the derivatives. Let us consider the possible index structures and their weights in turn. Since the barred derivative only lives in the first $d$ dimensions, we find that
\begin{equation}
	 \HH_{{\cal M} \overline{\cal R}} \partial^{\overline{\cal R}} \sim \XX_{{\cal M} \bar{\mu}}\partial^{\bar{\mu}}  + \ZZ_{{\cal M} \bar{\mu}} \partial^{\bar{\mu}} .
\end{equation}
This implies that the extended graviton field contracting with the derivative has weight
\eq{
	w \! \left[ \HH_{{\cal M} \overline{\cal R}}\partial^{\overline{\cal R}} \right] =0, +2.
	\label{eq:weight_dH}
}
On the other hand, the nilpotency of $\XX_\MM$ and $\ZZ_\MM$ implies that
\begin{equation}
	\HH_{{\cal M} \overline{{\cal R}}}\HH^{{\cal N} \overline{\cal R}} \sim X_{\mu \bar{\rho}}Z^{\nu \bar{\rho}} + Z_{\mu \bar{\rho}}X^{\nu \bar{\rho}} + Y_{\mu}Y^{\nu},
\end{equation}
which in turn fixes the weight
\begin{equation}
w \! \left[\HH_{{\cal M} \overline{{\cal R}}}\HH^{{\cal N} \overline{\cal R}}\right] = +2.\label{eq:weight_HH}
\end{equation}
From \Eqs{eq:weight_dH}{eq:weight_HH}, we conclude that the weights for cubic and quartic order are
\be 
	w \! \left[ \mathcal{L}^{(3)}_{\textrm{G}} \right] \ge +2 \qquad \text{and}\qquad w\! \left[ \mathcal{L}^{(4)}_{\textrm{G}} \right] \ge +2.
\ee
Because extended gravity is a two-derivative theory, all higher-order interaction terms  share the same derivative structures as \Eq{eq:Sfat_schematic} except with more powers of the extended graviton. Since $w \! \left[\HH_{{\cal M} \overline{{\cal R}}}\HH^{{\cal N} \overline{\cal R}}\right] >0$, terms at quintic order and higher have $w > +2$ and can thus be dropped.

\subsubsection*{Transverse Conditions}

As before, we can exploit the transverse properties of the fields to eliminate even more terms.
Up to total derivatives, the action in \Eq{eq:LagrBI} is equal to
\be
\begin{aligned}
\LL^{(3)}_{\rm BI} = &\phantom{+}\frac{1}{2\sqrt{2}} Z^{\mu \bar \nu} \partial_\mu Y_{\rho} \partial_{\bar \nu} Y^{\rho}
+ \frac{1}{2\sqrt{2}} Z^{\mu \bar \nu} \partial_{\bar\nu} Y_{\mu} \partial_{\rho} Y^{\rho} 
- \frac{1}{\sqrt{2}} Z^{\mu \bar \nu}\partial_{\rho} Y_{\mu} \partial_{\bar\nu} Y^{\rho}\\&+\frac{1}{2\sqrt{2}}Z^{\mu\bar\nu}\partial_\mu Z^{\rho\bar\sigma} X_{\rho\bar{\nu}\bar{\sigma}} - \frac{1}{2\sqrt{2}}Z^{\mu\bar\nu}\partial^\rho Z_\mu^{\;\;\bar\sigma} X_{\rho\bar\nu\bar\sigma} - \frac{1}{2\sqrt{2}}Z^{\mu\bar\nu}Z^{\rho\bar\sigma} \partial_\mu X_{\rho\bar\nu\bar\sigma} +{\cal O}(\partial_{\bar\nu}Z^{\mu\bar\nu}) \\
\LL^{(4)}_{\rm BI} =& - \frac{1}{16} Z^{\mu \bar \nu} Z_\mu^{\;\; \bar \rho} \partial_{\bar \nu} Y_\sigma \partial_{\bar \rho} Y^\sigma 
- \frac{1}{16} Z^{\mu \bar \nu} Z^{\rho \bar\sigma} \partial_{\bar \nu} Y_\mu \partial_{\bar \sigma} Y_\rho
+ \frac{1}{8} Z^{\mu \bar \nu} Z^{\rho \bar\sigma} \partial_{\bar \nu} Y_\rho \partial_{\bar\sigma} Y_\mu \\& +\frac{1}{8} Z^{\mu\bar\nu}Z_{\mu\bar\alpha}\partial_{\bar\nu}Z_{\rho\bar\sigma} X^{\rho\bar\sigma\bar\alpha} +\frac{1}{8}Z^{\mu\bar\nu}Z^{\rho\bar\sigma} \partial_{\bar\sigma} Z_\mu^{\;\;\bar\alpha} X_{\rho\bar\nu\bar\alpha}  + {\cal O}(\partial_{\bar\nu}Z^{\mu\bar\nu}),
\end{aligned}
\label{eq:BI_transmutation}
\ee
where we have defined the right-index field strength for $X_{\mu\bar\nu}$,  
\be
X_{\mu\bar \nu\bar \rho} = \partial_{\bar \nu} X_{\mu \bar\rho}-\partial_{\bar \rho} X_{\mu \bar\nu}.
\ee
Crucially, the field $X_{\mu\bar \nu}$ only appears in the action through its field strength $X_{\mu\bar\nu \bar\rho}$ or in terms proportional to $\partial_{\bar\nu}Z^{\mu\bar\nu}$. This allows us to apply an argument similar to that in \Sec{sec:YM}.    In particular, any factor of $\partial_{\bar\nu}Z^{\mu\bar\nu}$ that contributes to a Feynman diagram will ultimately be projected to zero on an external leg or attached to an internal vertex.  Since all internal vertices involving $X_{\mu\bar\nu}$ depend only on the field strength $X_{\mu\bar\nu \bar\rho}$ or are proportional to $\partial_{\bar\nu}Z^{\mu\bar\nu}$, these longitudinal contributions are always eventually zeroed out. The resulting BI action in \Eq{eq:BI_transmutation} also agrees with an action-level double copy construction combining YM theory and the NLSM~\cite{futurework}.

Let us contrast the quartic representation of BI action in \Eq{eq:BI_transmutation} with the canonical representation of BI action arising from brane-localized gauge fields,
\be
{\cal L}_{\rm BI} = -T \sqrt{-{\rm det}(\eta_{\mu\nu} + 2\pi \alpha' F_{\mu\nu})}, 
\ee
where the determinant structure induces an infinite tower of interactions.  As before, all of the dimensionful coupling constants in our new BI action are absorbed into the normalization of the longitudinal polarizations. Similar to the NLSM, our formulation does not manifest permutation invariance and unitarity, though these are still present in scattering amplitudes, as proved in \Ref{unified}. However, thanks to its finite interactions, it is tremendously simpler to calculate amplitudes in this action. In \Ref{Rocek:1997hi}, a simplification of the BI action was constructed using auxiliary fields and a setup specific to certain spacetime dimensions. In contrast, our formulation is valid in arbitrary spacetime dimension and the construction follows directly from our analogous treatment of the NLSM; it may therefore also offer some insight for the double copy structure of BI theory.

\subsection{Dimensional Reduction to the Special Galileon Theory}

Last but not least, we apply action-level transmutation again to BI theory to obtain an action for the SG theory.  This is equivalent to a double dimensional reduction of the extended graviton action, taking the unbarred and barred indices of the extended graviton to both run over $2d+1$ dimensions.  We then decompose the extended graviton field $\HH_{\MM \overline \MM}$ into a $(2d+1) \times (2d+1)$ matrix,
\be 
\HH_{\MM \overline \MM}  = \XX_{\MM \overline \MM}+\YY_{\MM \overline\MM}+\ZZ_{\MM\overline\MM},
\ee
where each contribution is
\be 
\begin{aligned}
\XX_{\MM \overline \MM} &= \frac{1}{2}\left(
\begin{array}{ccc}
 X_{\mu \bar\mu}  & 0 &   - i X_{\mu \bar\mu}  \\
0 & 0  &  0 \\
-i  X_{\mu \bar\mu}  & 0 &  -  X_{\mu  \bar\mu} \end{array}
\right)\\
\YY_{\MM \overline \MM} &=  \left(
\begin{array}{ccc}
0 & 0 &  0 \\
0 & Y  &  0 \\
0 & 0 &  0
\end{array}
\right)\\
\ZZ_{\MM \overline \MM} &= \frac{1}{2} \left(
\begin{array}{ccc}
 Z_{\mu \bar\mu}  & 0 &   +i Z_{\mu \bar\mu}   \\
0 & 0  &  0 \\
+i Z_{\mu \bar\mu}   & 0 &  -Z_{\mu \bar\mu} 
\end{array}
\right). \label{eq:HHblock}
\end{aligned}
\ee
The components not shown all enter in pairs in the action so they can be consistently dropped from the action provided we are interested in tree-level amplitudes only involving the states represented above. These definitions again imply a nilpotency condition,
\be 
\begin{aligned}
&\XX_{\MM \overline \RR} \XX_\NN^{\;\;\overline \RR} \!\!\!&= \ZZ_{\MM \overline \RR} \ZZ_\NN^{\;\; \overline \RR}&= \XX_{\RR\overline\MM}\XX^\RR_{\;\;\overline \NN} \!\!\!&= \ZZ_{\RR\overline\MM}\ZZ^\RR_{\;\;\overline \NN} &=0 \\
& \XX_{\MM \overline \RR} \YY_\NN^{\;\;\overline \RR}\!\!\!&=\ZZ_{\MM \overline \RR} \YY_\NN^{\;\; \overline \RR}  &= \XX_{\RR\overline\MM}\YY^\RR_{\;\;\overline \NN} \!\!\!&= \ZZ_{\RR\overline\MM}\YY^\RR_{\;\;\overline \NN}
&=0 ,
\end{aligned}
\ee 
and likewise for structures with additional derivatives. 
Using the weight-counting arguments provided at the end of this section, we can expand the extended gravity action in components and truncate, yielding the action for the SG,
\be 
\LL_{\rm SG} =  \LL^{(2)}_{\rm SG} + \LL^{(3)}_{\rm SG} ,
\ee
where each term is given by 
\be 
\begin{aligned}
\LL^{(2)}_{\rm SG} =&\phantom{+}   X_{\mu \bar\mu} \Box Z^{\mu\bar \mu}  + \frac{1}{2} Y \Box Y  \\
\LL^{(3)}_{\rm SG}  =&  
-\frac{1}{2} Z^{\mu \bar \mu} \partial_{\bar \mu} Z^{\nu \bar\nu} \partial_\nu  X_{\mu \bar\nu}
+ \frac{1}{4} Z^{\mu \bar \mu} \partial_\mu Y \partial_{\bar \mu} Y  + \{ YY \rightarrow X_{\mu\bar\mu} Z^{\mu\bar\mu} + Z_{\mu\bar\mu} X^{\mu\bar\mu} \}  \\&+ {\cal O}(\partial_\mu Z^{\mu\bar\mu}) + {\cal O}(\partial_{\bar \mu} Z^{\mu\bar\mu}).
\end{aligned}
\ee
As we will show, all possible quartic interactions in the extended graviton action can be consistently dropped due to the weight counting argument presented in the subsequent discussion.  Moreover, all quintic and higher-order interactions can also be dropped because the SG action is equivalent to a transmutation of BI action, which itself originates from the extended graviton action truncated to quartic order.

\subsubsection*{Weight Counting}

From \Eqs{eq:Gal_from_BI}{eq:BI_amps}, we see that the SG amplitude is given by
\be 
	A(\phi_1, \phi_2, \ldots , \phi_{n-1}, \phi_n) = A(Y_1,Z_{2}, \ldots, Z_{n-1}, Y_n),
	\label{eq:SG_amps}
\ee
corresponding to a pair of $Y$ states with all other external states given by $Z_{\mu\bar\mu}$ particles whose polarizations are longitudinal and thus proportional to $p_\mu p_{\bar\mu}$. Note that these external states are the ``square'' of the $Y$ and $Z_{\mu}$ external states for the NLSM in Eq.~\eqref{eq:NLSM_amp}.

Since dimensional reduction is applied to both barred and unbarred indices, it is natural to promote the weight into a two-component vector, $(w, \bar w) = (n-h, n-\bar h)$, where $h$ and $\bar h$ correspond to the pseudo-helicity for the unbarred and barred indices. For each state, we have
\be 
	w[X_{\mu\bar\mu}] = 	\bar w[X_{\mu\bar\mu}]  = +2, \qquad
	w[Y] = \bar w[Y] = +1 ,  \qquad 
	w[Z_{\mu\bar\mu}] = \bar w [Z_{\mu\bar\mu}] = 0.
	\label{eq:field_weight_vec}
\ee
For an amplitude on its factorization channel, $A\sim A_L A_R$, their weights are related by $w[A] = w[A_L]+w[A_R]-2$ and $\bar w[A] = \bar w[A_L]+\bar w[A_R]-2$, so we conclude that the tree-level SG amplitude has weight $(w,
\bar w) = (+2,+2)$.
From the schematic form in Eq.~\eqref{eq:Sfat_schematic}, the unbarred indices in $\HH^4  \bar{\partial} \bar{\partial}$ have the same tensor structure as the quartic interactions in YM. As we learned in the NLSM, these interactions have $w =+4$, which can be truncated, and similarly for $\HH^4 {\partial}{\partial}$.
Combining with the weight counting arguments in BI, we see that the quartic and higher interactions of the extended graviton action dimensionally reduce to terms with either $w > +2$ or $\bar w >+2$, so they can be consistently dropped.  

\subsubsection*{Transverse Conditions}

Next, let us consider the transverse properties of the fields in the SG action.
Defining an analogue of the Riemann tensor as in \Ref{XYZ}, 
\be
X_{\mu\nu\bar{\mu}\bar{\nu}} = \partial_\mu \partial_{\bar\mu}X_{\nu\bar\nu} + \partial_\nu \partial_{\bar\nu} X_{\mu\bar\mu} - \partial_\mu \partial_{\bar\nu} X_{\nu\bar\mu} - \partial_\nu \partial_{\bar\mu} X_{\mu\bar\nu} = \partial_\mu X_{\nu\bar\mu\bar\nu} - \partial_\nu X_{\mu\bar\mu\bar\nu}, 
\ee
our final form for the SG action becomes
\be 
\LL_{\rm SG} =X_{\mu \bar\mu} \Box Z^{\mu\bar \mu}  + \frac{1}{2} Y \Box Y  -\frac{1}{4}\left(X_{\mu\nu\bar\mu\bar\nu} Z^{\mu\bar\mu}Z^{\nu\bar\nu} + Z^{\mu\bar\nu}Y {\partial}_\mu {\partial}_{\bar\nu} Y\right) + {\cal O}(\partial_\mu Z^{\mu\bar\mu}) + {\cal O}(\partial_{\bar \mu} Z^{\mu\bar\mu}),\label{eq:SGfinal}
\ee
modulo total derivatives.
Up to terms of the form ${\cal O}(\partial_\mu Z^{\mu\bar\mu})$ and ${\cal O}(\partial_{\bar \mu} Z^{\mu\bar\mu})$, the field $X_{\mu\bar \mu}$ appears in the action only in the form of $X_{\mu\nu\bar{\mu}\bar{\nu}}$. By an argument exactly analogous to the one given in Sec.~\ref{sec:BI}, these terms proportional to ${\cal O}(\partial_\mu Z^{\mu\bar\mu})$ and ${\cal O}(\partial_{\bar \mu} Z^{\mu\bar\mu})$ can be dropped. As was shown in \Ref{XYZ}, \Eq{eq:SGfinal} can also be obtained from \Eq{eq:LNLSM} via the action-level double copy.

The cubic SG action in \Eq{eq:SGfinal} is substantially simpler than the canonical formulation of the SG action, which describes a scalar invariant under an extended shift symmetry \cite{kurt},
\eq{
\phi \rightarrow \phi + a + b_\mu + c_{\mu\nu}x^\mu x^\nu +  c^{\mu\nu} \partial_\mu \phi \partial_\nu \phi/\Lambda^6,
}
where $a$, $b_\mu$, and $c_{\mu\nu}$ are a constant scalar, vector, and traceless symmetric tensor, respectively.  In four dimensions, the canonical form of the SG action is 
\be
\mathcal{L}_{\rm SG} = -\frac{1}{2} \partial_\mu \phi \partial^\mu \phi + \frac{1}{12\Lambda^6}  \partial_\mu \phi \partial^\mu \phi  \left( \Box \phi \Box \phi  - \partial_{\rho} \partial_\sigma\phi  \partial^{\rho} \partial^{\sigma} \phi 
\right),
\ee
while in $d$ dimensions there is a tower of even-point interactions at all valences less than or equal to $d+1$.  In contrast, the SG action in \Eq{eq:SGfinal} is purely cubic for general dimension.

\subsection{Origin of the Double Copy}

By inspection, the SG action in \Eq{eq:SGfinal} is obtained by squaring all of the terms in the NLSM action in \Eq{eq:XYZ}. This is an action-level manifestation of the double copy \cite{XYZ}. Our prescription for dimensional reduction actually trivializes the origin of the double copy structure, by the following argument.  Our discussion of weight counting reveals that pion scattering is encoded within the cubic sector of YM theory.  However, the purely cubic topologies of YM theory automatically satisfy kinematic Jacobi identities up to contact terms coming from the quartic vertices.  This is because one can always probe a maximal factorization channel on which every propagator is on shell.  In this limit, the only contributions to amplitudes are the cubic diagrams, so these  contributions necessarily satisfy the kinematic Jacobi identities up to terms involving contact terms.  However, since all quartic terms are eliminated by the choice of external states corresponding to pion scattering, the mismatch from the kinematic Jacobi identities is eliminated and the resulting cubic action automatically satisfies them.  We thus conclude that  since the cubic sector of YM double copies into the cubic sector of gravity and these coincide with the NLSM and the SG, the actions that result from our dimensional reduction automatically manifest the double copy.

\section{Conclusions}
\label{sec:conclusions}

In this paper, we have proposed a variation of dimensional reduction that excises the NLSM from YM theory as well as BI theory and the SG theory from the extended graviton action.  This operation is essentially an action-level incarnation of the transmutation operation on scattering amplitudes derived in \Ref{unified}.   These relations reveal the origin of the kinematic algebra of the NLSM as the higher-dimensional Poincar\'e invariance of an underlying YM theory.   Remarkably, the NLSM and SG arise from purely cubic interactions in YM and gravity, while BI arises from only the cubic and quartic interactions of gravity.  Since the cubic sector of YM theory automatically double copies into gravity, the same is trivially true for the NLSM to the SG. Note that the theories obtained here---the NLSM, BI theory, and the SG theory---precisely coincide with the exceptional theories studied in \Ref{Cheung:2016drk} argued to be the natural effective field theory analogues of YM theory and gravity.

Our results suggest a number of directions for future work.  One avenue is to derive action-level versions of the other transmutation operations presented in \Ref{unified}.  For instance, one expects an action-level operation that sends gravity to YM theory.  While this is naturally accomplished by Kaluza-Klein reduction, the simplicity of the S-matrix mapping suggests that something more minimal is possible. Such a realization may teach us new structures of YM theory, such as color-kinematics duality. 

Another direction deserving of further study is higher loop order in perturbation theory.  Since \Ref{unified} derived unifying relations for tree-level scattering amplitudes, the procedure for dimensional reduction derived here is only guaranteed to reproduce amplitudes at tree level.  As is also the case for the double copy construction, matching at higher loop order will likely involve additional structure.   It would also be interesting to study the loop-level amplitudes computed from the actions presented here and to compare them with known results in the NLSM, BI theory, and the SG theory.

Last but not least, pions are famously known to be related to gluons through the Goldstone boson equivalence theorem. 
Although the $(2d+1)$-dimensional transmutation is proven in \Ref{unified} by modern S-matrix techniques, 
it would be illuminating to show the connection to the Goldstone boson equivalence theorem. 
Such a relation would also offer new insights into the nature of transmutation.

\begin{center} 
 {\bf Acknowledgments}
 \end{center}
 \noindent 
We thank Andrés Luna, John Joseph M.~Carrasco, Song He, and Yu-tin~Huang for helpful discussions. C.C. is supported by a Sloan Research Fellowship and C.C., C.-H.S., and C.W. are supported in part by a DOE Early Career Award under Grant No. DE-SC0010255 and by the NSF under Grant No. NSF PHY-1125915. G.N.R.~was supported at Caltech by a Hertz Graduate Fellowship and a NSF Graduate Research Fellowship under Grant No.~DGE-1144469 and is currently supported at University of California, Berkeley by the Miller Institute for Basic Research in Science. C.-H.S. is also supported by Mani L. Bhaumik Institute for Theoretical Physics.  This material is based upon work supported by the U.S. Department of Energy, Office of Science, Office of High Energy Physics, under Award Number DE-SC0011632.

\bibliographystyle{utphys-modified}
\bibliography{gravity_action}

\end{document}